\documentclass[10pt,aps,prb,twocolumn,showpacs,amsmath,amssymb,superscriptaddress]{revtex4-1}
\usepackage{graphicx}
\usepackage{natbib}
\usepackage{amsmath}
\usepackage{xcolor}
\usepackage{comment}
\usepackage{soul}
\usepackage[colorlinks=true, urlcolor  = blue, citecolor = blue, linkcolor = blue]{hyperref}
\usepackage{siunitx}

\begin{document}

\title[The magnetic ground state of Cu$_{2}$OSeO$_{3}$ from long-wavelength neutron diffraction]{Investigating the magnetic ground state of the skyrmion host material Cu$_{2}$OSeO$_{3}$ using long-wavelength neutron diffraction}

\author{K\'{e}vin J. A. Franke}
\affiliation{Durham University, Centre for Materials Physics, Durham, DH1 3LE, United Kingdom}

\author{Philip R. Dean}
\affiliation{Durham University, Centre for Materials Physics, Durham, DH1 3LE, United Kingdom}

\author{Monica Ciomaga Hatnean}
\affiliation{University of Warwick, Department of Physics, Coventry, CV4 7AL, United Kingdom}

\author{Max T. Birch}
\affiliation{Durham University, Centre for Materials Physics, Durham, DH1 3LE, United Kingdom}

\author{Dmitry Khalyavin}
\affiliation{ISIS Facility, STFC Rutherford Appleton Laboratory, Chilton, Didcot, Oxfordshire, OX11 0QX, United Kingdom}

\author{Pascal Manuel}
\affiliation{ISIS Facility, STFC Rutherford Appleton Laboratory, Chilton, Didcot, Oxfordshire, OX11 0QX, United Kingdom}

\author{Tom Lancaster}
\affiliation{Durham University, Centre for Materials Physics, Durham, DH1 3LE, United Kingdom}

\author{Geetha Balakrishnan}
\affiliation{University of Warwick, Department of Physics, Coventry, CV4 7AL, United Kingdom}

\author{Peter D. Hatton}
\affiliation{Durham University, Centre for Materials Physics, Durham, DH1 3LE, United Kingdom}

\date{\today}

\begin{abstract}
We present long-wavelength neutron diffraction data measured on both single crystal and polycrystalline samples of the skyrmion host material Cu$_{2}$OSeO$_{3}$. We observe magnetic satellites around the $(0\bar{1}1)$ diffraction peak not accessible to other techniques, and distinguish helical from conical spin textures in reciprocal space. We confirm successive transitions from helical to conical to field polarised ordered spin textures as the external magnetic field is increased. 
The formation of a skyrmion lattice with propagation vectors perpendicular to the field direction is observed in a region of the field-temperature phase diagram that is consistent with previous reports. 
Our measurements show that not only the field-polarised phase but also the helical ground state are made up of ferrimagnetic clusters instead of individual spins. These clusters are distorted Cu tetrahedra, where the spin on one Cu ion is anti-aligned with the spin on the three other Cu ions.
\end{abstract}

\maketitle

\section{Introduction}
Skyrmions are topologically protected, nano-sized swirls of spins found in a range of magnetic materials. In recent years a number of spectacular advances have demonstrated the existence, not only of magnetic skyrmions, but also their ordering into a skyrmion lattice (SL) in several magnetic materials with a chiral structure.\cite{muhlbauer_skyrmion_2009, yu_real-space_2010, yu_near_2011, seki_observation_2012, tokunaga_new_2015, kezsmarki_neel-type_2015, bordacs_equilibrium_2017, kurumaji_neel-type_2017} This symmetry can lead to the formation of an antisymmetric Dzyaloshinskii-Moriya interaction (DMI)\cite{dzyaloshinsky_thermodynamic_1958, moriya_anisotropic_1960} that favours canting between neighbouring spins. Competition between the symmetric exchange interaction promoting parallel alignment of spins and the DMI leads to a magnetic ground state that generally consists of helices, long period magnetic structures where the magnetisation rotates in a plane perpendicular to the propagation direction. On the application of a magnetic field a conical (C) structure forms, with a net magnetic moment along the propagation direction. The C structure exists over a wide range of applied magnetic fields with the cone angle decreasing upon increasing the field until it is reduced to zero and magnetic moments align with the field direction. Skyrmions are very close in energy to the C structure, and the SL phase is generally stabilised in a small temperature and field region of the phase diagram just below the critical temperature $T_{\textrm{c}}$. Theoretical predictions show that for skyrmions to exist systems require an easy-axis ferromagnetism on top of a DMI.\cite{bogdanov_thermodynamically_1989,bogdanov_thermodynamically_1994}

The phase diagram of Cu$_{2}$OSeO$_{3}$ resembles that of other skyrmion hosting systems such as several B20 compounds,\cite{yu_real-space_2010, yu_near_2011} or $\beta$-Mn-type CoZnMn alloys \cite{tokunaga_new_2015}: It consists of a helical (H) ground state, a C phase in applied fields, a field polarised (FP) phase in large applied fields, and a SL phase in a pocket in applied field and temperature just below $T_{\textrm{c}}$.\cite{seki_observation_2012}
The helical ground state was determined by Lorentz TEM.\cite{seki_observation_2012, adams_long-wavelength_2012, seki_formation_2012, yang_strong_2012} Earlier neutron powder diffraction was unable to determine this structure due to the long helical wavelength of $\sim\!65$ nm ($Q \sim 0.0015$ \AA$^{-1}$). \cite{bos_magnetoelectric_2008,adams_long-wavelength_2012,seki_formation_2012} The phase diagram determined from Lorentz TEM has been confirmed by small angle neutron scattering (SANS) which finds that the ground state helices are oriented along $\{100\}$ directions, but cannot easily differentiate between a cone and a helix. \cite{adams_long-wavelength_2012,seki_formation_2012} In SANS, the formation of a SL is observed as a sixfold diffraction pattern, consistent with the hexagonal close packing of skyrmions in a plane perpendicular to the applied field direction.

Previous investigations have either shown that Cu$_{2}$OSeO$_{3}$ is a ferrimagnet \cite{kohn_new_1977, bos_magnetoelectric_2008, belesi_ferrimagnetism_2010} without addressing the chiral nature of the magnetic ground state, or investigated the chirality without taking into account ferrimagnetism.\cite{seki_observation_2012, adams_long-wavelength_2012, seki_formation_2012, yang_strong_2012, zhang_direct_2017} Here we investigate the magnetic microstructure of the magnetic phases of Cu$_{2}$OSeO$_{3}$ and address the question of whether the ground state H structure is also ferrimagnetic. 
Using time-of-flight long-wavelength neutron diffraction on a single crystal sample, we observe magnetic satellites around the $(0\bar{1}1)$ diffraction peak not accessible to other techniques. Furthermore, our measurements uniquely allow us to distinguish helical from conical spin textures in reciprocal space. We show successive transitions from H to C to FP as the external magnetic field is increased in the magnetically ordered phases, and the formation of a skyrmion lattice with propagation vectors perpendicular to the field direction in a region of the field-temperature phase diagram that is consistent with previous reports.\cite{seki_observation_2012, adams_long-wavelength_2012}
We use powder diffraction measurements to determine the magnetic microstructure of the FP and H phases, and show that not only the FP phase but also the H ground state are made up of ferrimagnetic clusters instead of individual spins. 

\section{Methods}
Polycrystalline samples of Cu$_{2}$OSeO$_{3}$ were synthesized as described previously.\cite{bos_magnetoelectric_2008} Phase purity was confirmed by powder x-ray diffraction. The same polycrystalline powder was also used for the growth of single crystals by the chemical vapour transport technique following the procedure described by \citet{seki_observation_2012}

DC magnetisation measurements were performed using SQUID magnetometry, in a Quantum Design MPMS. AC susceptibility measurements were performed on the same instrument at an excitation frequency of $10$ Hz with an amplitude of $0.1$ mT.

Neutron powder and single crystal diffraction experiments were performed on the time-of-flight long-wavelength neutron diffractometer WISH \cite{chapon_wish:_2011} at the ISIS Facility of the Rutherford Appleton Laboratory (UK) to determine the crystal and magnetic structures. 
The WISH detector system consists of pixelated $^{3}$He gas tubes covering scattering angles from $2\Theta =10^{\circ}$ to $2\Theta =170^{\circ}$ in the plane and $\pm 12.8^{\circ}$ out of the plane. Powder diffraction patterns are measured at fixed scattering angles $2\Theta$ as a function of  the time-of-flight (which is related to $d$-spacing).
The highest resolution is offered at large scattering angles which require longer wavelength neutrons. This, however, limits the highest observable $d$-spacing.
For measuring powder diffraction patterns, we thus consider two detector banks, at $2\Theta =27^{\circ}$ and $2\Theta =58^{\circ}$. For single crystal diffraction patterns we restricted the neutron wavelength to $8.2\pm0.2$ \AA ~to observe the $(0\bar{1}1)$ Bragg peak and its magnetic satellites. 

Data reduction, analysis, and simulation of single crystal diffraction patterns was done using the Mantid software.\cite{Mantid2014} Rietveld refinement of powder diffraction data was undertaken using the FullProf suite of programs.\cite{FullProf1993}

\section{Results and Discussion}

\begin{figure}
\centering
\includegraphics[width=1\columnwidth]{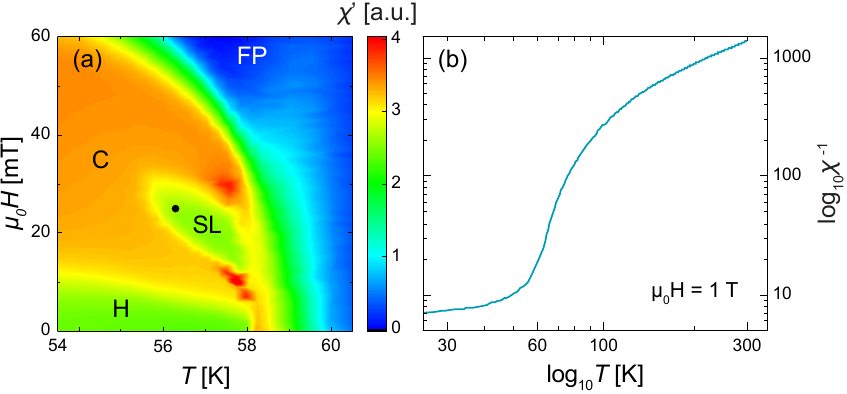}
\caption{\label{MPMS} (a) AC susceptibility mapping of the phase diagram of Cu$_{2}$OSeO$_{3}$ for H~$\parallel[111]$. The black dot indicates where the single crystal diffraction data in Figure~\ref{single-crystal}(e) was measured. (b) Inverse susceptibility as a function of temperature.}
\end{figure}

\paragraph*{Magnetisation:}
AC susceptibility measurements have proven powerful in determining the presence and location of the SL.\cite{bauer_magnetic_2012, bauer_generic_2016} Figure~\ref{MPMS}(a) shows the real part $\chi'$ of the AC susceptibility from measurements on Cu$_{2}$OSeO$_{3}$ with the magnetic field $\mu_{0}H$ applied along the $[111]$ direction. The SL is identified as a decrease in $\chi'$ relative to the surrounding C phase. It is located in the range $12\leq \mu_{0}H \leq30$~mT and $55.75\leq T\leq 58$~K in agreement with previous observations.\cite{seki_observation_2012, adams_long-wavelength_2012}

The inverse (DC) susceptibility $1/\chi$ presented in Figure~\ref{MPMS}(b) confirms a critical temperature $T_{\textrm{c}}\approx59$~K. Above $T_{\textrm{c}}$, however, the temperature dependence of $1/\chi$ does not agree with the earlier report by \citet{bos_magnetoelectric_2008}:  in the PM phase, we observe a negative curvature (noted in an early paper by \citet{kohn_new_1977}), instead of a linear increase with temperature as observed for ferromagnets and antiferromagnets. This dependence is characteristic of ferrimagnets,\cite{kittel_introduction_1996} and is particularly clear with the log-log scale used here. Magnetisation measurements in the PM phase thus indicate a ferrimagnetic alignment of spins.

\begin{figure*}
\centering
\includegraphics[width=1\textwidth]{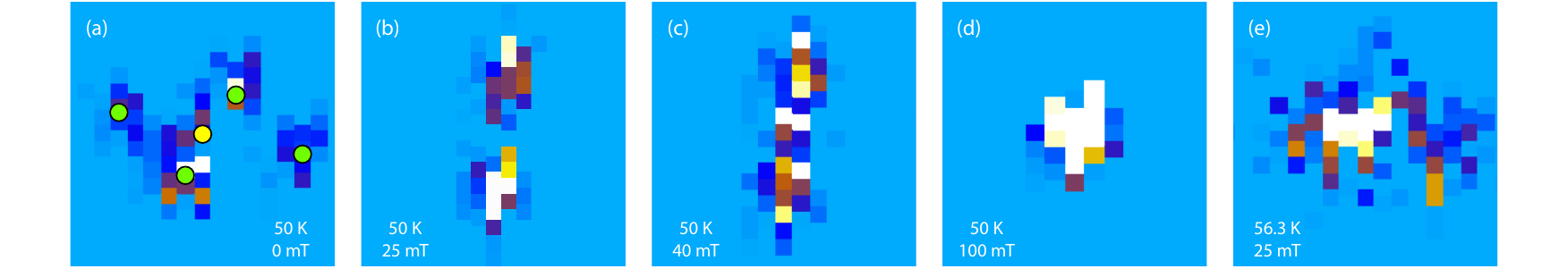}
\caption{\label{single-crystal} (a)-(d) Magnetic contribution to single crystal neutron diffraction data measured around the $(0\bar{1}1)$ Bragg peak (yellow dot) at $50$ K as a function of increasing applied magnetic field. Green dot are fits to the helical peaks. (e) Measurement in the SL phase [c.f. black dot in Figure~\ref{MPMS}(a)].}
\end{figure*}

\paragraph*{Single Crystal Diffraction:}
For single crystal diffraction the sample was also mounted with the applied field along the $[111]$ direction. Data measured around the $(0\bar{1}1)$ Bragg peak (yellow dot) is presented in Figure~\ref{single-crystal}. The structural peak is removed by subtracting the structural diffraction pattern measured in zero field in the PM phase, thus showing only magnetic peaks. 
The technique is distinct from SANS, where the scattering around the transmitted neutron beam is investigated. The operating principle of our single crystal neutron diffraction is more similar to resonant elastic x-ray scattering,\cite{langner_coupled_2014, zhang_resonant_2016} in that periodically modulated spin textures are observed as satellite peaks around a structural Bragg reflection. As resonant elastic x-ray scattering relies on matching the photon energy (and thus wavelength) to a x-ray absorption edge, only the $(001)$ peak of Cu$_{2}$OSeO$_{3}$ is accessible at the Cu $L_{3}$ edge (in many other skyrmion hosts, such as FeGe or MnSi, no Bragg peaks are accessible at all).\cite{zhang_resonant_2016} Single crystal neutron diffraction has the advantage of allowing access to a wide range of structural diffraction peaks, such as the $(0\bar{1}1)$ peak considered here. It furthermore allows distinguishing helices and cones as detailed below.

At $50$ K, in the absence of an applied magnetic field [Fig.~\ref{single-crystal}(a)], satellite peaks (green dots) oriented along $\{100\}$ directions are observed as expected in the H phase \cite{adams_long-wavelength_2012,seki_formation_2012}: Two peaks observed to the left and right of the Bragg peak correspond to the $(100)$ direction. Two peaks above and below the Bragg peak correspond to the $(010)$ and $(001)$ directions (these satellites coincide due to the orientation of the sample in the  instrument). We extract a helical $Q$ of $0.0016(3)$ \AA$^{-1}$, in agreement with previous SANS measurements.\cite{adams_long-wavelength_2012,seki_formation_2012} Upon the application of a magnetic field ($\mu_{0}H=25$ mT), the helices rotate towards the field direction [Fig.~\ref{single-crystal}(b)]. No central peak is observed indicating the absence of a net magnetic moment as expected for a helix. A further increase of the magnetic field to $\mu_{0}H=40$ mT [Fig.~\ref{single-crystal}(c)] induces a spin canting and the structure becomes conical. This is observed as the appearance of intensity at the parent peak position indicating a net magnetic moment. When the applied field is increased further [Fig.~\ref{single-crystal}(d)], all moments align with the field ($\mu_{0}H=100$ mT). Figure~\ref{single-crystal}(e) shows the diffraction pattern measured at $56.3$~K in $25$~mT, at the location of the skyrmion phase [c.f. black dot in Figure~\ref{MPMS}(a)]. The resolution is too low to distinguish separate peaks, but a transfer of intensity from aligned along the applied field direction to the plane perpendicular to the applied field is observed, showing that the propagation vectors of the skyrmion lattice are in a plane perpendicular to the applied field. The lack of magnetic satellites above and below the parent Bragg peak precludes the coexistence of the skyrmion phase with the C phase in our sample. 

The resolution of our single crystal diffraction data is not sufficient to easily distinguish the 6-fold SL satellites. However, it shows that the magnetic phase can be transformed from H to FP via the C phase by an increase in the applied magnetic field $\mu_{0}H$, and to the SL phase by a change in temperature.  Single crystal neutron diffraction measurements agree with the phase diagram established by AC susceptibility measurements and the helical wavelength established in SANS measurements.\cite{adams_long-wavelength_2012,seki_formation_2012}

\begin{figure}[b]
\centering
\includegraphics[width=1\columnwidth]{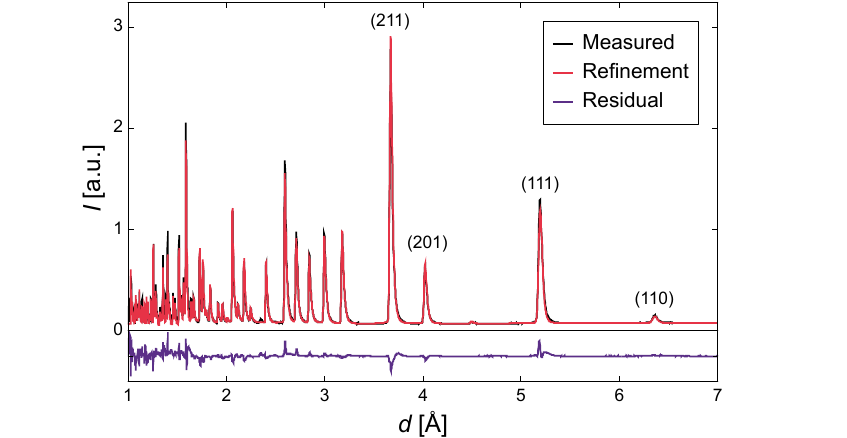}
\caption{\label{60K-ZF-90deg} Measured (black) neutron powder diffraction data of Cu$_{2}$OSeO$_{3}$ at $60$ K in zero applied magnetic field. Crystallographic parameters obtained by Rietveld refinement (red) are reported in Table~\ref{atompara}. Residuals (violet) have been offset for clarity.}
\end{figure}

\begin{figure}[b]
\centering
\includegraphics[width=1\columnwidth]{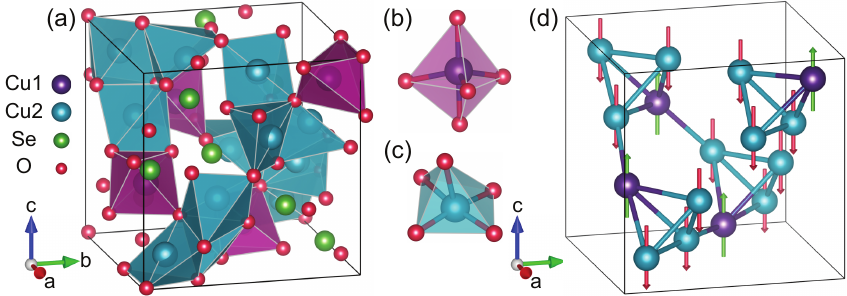}
\caption{\label{Structure} (a) Crystal structure of Cu$_{2}$OSeO$_{3}$ consisting of (b) Cu1 bipyramids and (c) Cu2 distorted square based pyramids. Bonds to Se ions have been omitted for clarity. (d) Representation of the ferrimagnetic structure with spins (green arrows) on Cu1 site antiparallel to the spins (red arrows) on Cu2 sites.}
\end{figure}

\paragraph*{Powder Diffraction:}
Neutron powder diffraction data measured at $60$ K in the PM phase is presented in Figure~\ref{60K-ZF-90deg}. The pattern was measured at zero magnetic field and thus contains only structural information. 
The crystallographic parameters obtained from Rietveld refinement are reported in Table~\ref{atompara}. The data is best fitted to the space group $P2_{1}3$ with lattice parameter $a=8.97639(7)$~\AA ~in agreement with previous reports.\cite{bos_magnetoelectric_2008}
The crystal structure of Cu$_{2}$OSeO$_{3}$ is shown in Figure~\ref{Structure}(a). The sixteen Cu ions in the unit cell sit on two different Wyckoff sites, four Cu1 ions on the 4a and twelve Cu2 ions on the 12b site. Cu1 ions bond with oxygen ions to form bipyramidal CuO$_{5}$ units, while Cu2 ions are bonded in distorted square based CuO$_{5}$ pyramids. The Cu$^{2+}$ ions are arranged in a network of distorted tetrahedra consisting of one Cu1 and three Cu2 ions.

\begin{table}
\centering
\footnotesize\rm
\caption{\label{atompara}Crystallographic parameters of Cu$_{2}$OSeO$_{3}$ at $60$ K obtained from Rietveld refinement of neutron powder diffraction data (Fig.~\ref{60K-ZF-90deg}). The space group is $P2_{1}3$ with lattice parameter $a=8.97639(7)$~\AA.}
\begin{tabular}{c c c c c}
\hline
Atom & Wyckoff site & $x$ & $y$ & $z$\\
\hline
Cu1	& $4a$	& $0.8870(8)$	& $0.8870(8)$	& $0.8870(8)$ \\ 	
Cu2	& $12b$	& $0.1315(7)$	& $0.1189(7)$	& $-0.1293(8)$ \\ 	
O1	& $4a$	& $0.7575(6)$	& $0.7575(6)$	& $0.7575(6)$ \\ 	
O2	& $4a$	& $0.018(1)$	& $0.018(1)$	& $0.018(1)$ \\ 	
Se1	& $4a$	& $0.2025(6)$	& $0.2025(6)$	& $0.2025(6)$ \\ 	
Se2	& $4a$	& $0.4636(5)$	& $0.4636(5)$	& $0.4636(5)$ \\ 	
O3	& $12b$	& $0.2686(6)$	& $0.1829(7)$	& $0.0324(7)$ \\ 	
O4	& $12b$	& $-0.0164(9)$& $0.0415(8)$	& $-0.2677(8)$ \\ 
\hline
\end{tabular}
\end{table}

\begin{figure}[b]
\centering
\includegraphics[width=1\columnwidth]{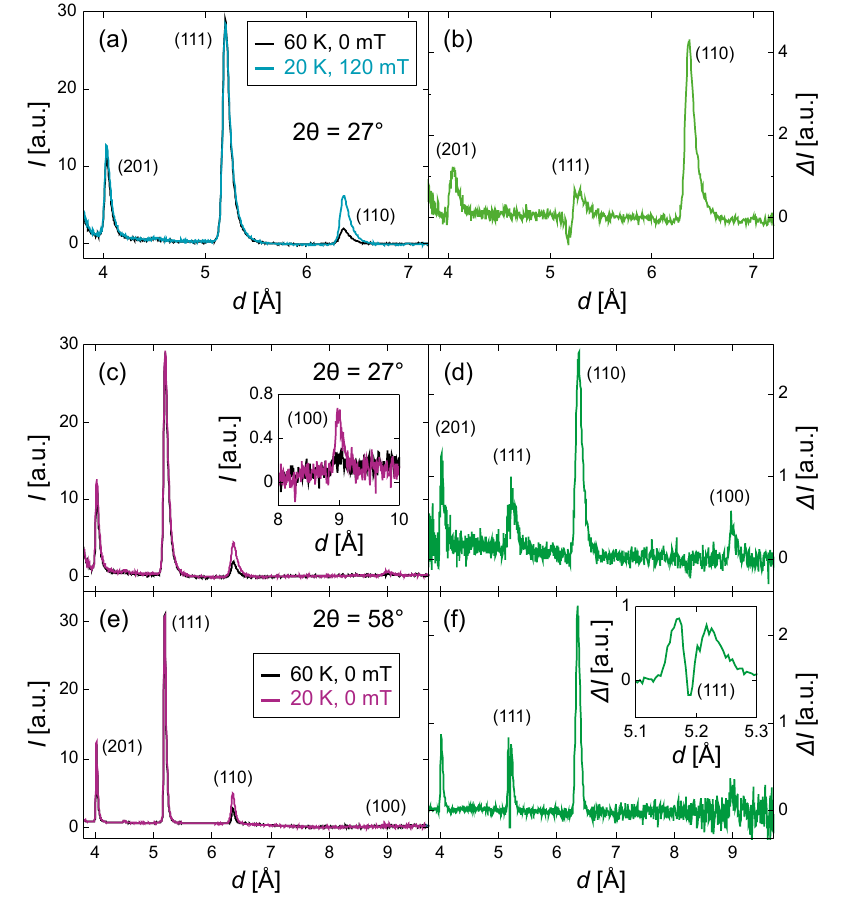}
\caption{\label{hel-FP-para} (a) Neutron powder diffraction data measured in the FP (blue) and PM (black) phases at a scattering angles of $2\Theta=27^{\circ}$. (b) Difference between both patterns. (c) \& (e) Data measured in the H (violet) and PM (black) phases at scattering angles of $2\Theta=27^{\circ}$ and  $2\Theta=58^{\circ}$. The difference between H and PM phases is shown in (d) and (f).}
\end{figure}

Data measured in the FP phase at $20$ K in an applied field $\mu_{0}H=120$ mT is shown in Figure~\ref{hel-FP-para}(a) for a scattering angle $2\Theta =27^{\circ}$. The diffraction pattern containing only structural information and measured in the PM phase is shown for comparison (black line). Analogous to our approach for single crystal diffraction, the difference $\Delta I$ between both patterns is shown in Figure~\ref{hel-FP-para}(b). $\Delta I$ should contain purely magnetic information, however, thermal contraction can lead to a small shift of structural peaks between the data measured at $60$~K and $20$~K. This can be identified by the shape of the difference peaks as a negative peak next to a positive peak (a shape resembling the derivative of a $\delta$-function). This is for example observed at the $(111)$ reflection in Figure~\ref{hel-FP-para}(b).
A clear magnetic contribution to the diffraction pattern is observed at the $(110)$ reflection. The data is best fitted with a model with spins on Cu1 ions aligned antiparallel to the spins on Cu2 ions, as sketched in Figure~\ref{Structure}(d), and a magnetic moment of $0.92(3)$ $\mu_{\text{B}}$ for each Cu ion. The FP phase is thus ferrimagnetic, consisting of clusters of four Cu ions, where the spin on one Cu1 ion is anti-aligned with the spins on three Cu2 ions.

The H phase was investigated at $20$ K in zero field. The data measured at $2\Theta =27^{\circ}$ and $2\Theta =58^{\circ}$ is again compared to the data measured in the PM phase in Figures~\ref{hel-FP-para}(c)-(f). In addition to the magnetic (110) peak already observed in the ferrimagnetic phase, additional purely magnetic peaks are observed at the $(100)$ position [as highlighted in the inset of Figure~\ref{hel-FP-para}(c)] and the $(111)$ position.  The higher resolution data at $2\Theta =58^{\circ}$ reveals the latter to be due to the presence of two magnetic satellite peaks, as shown in the inset of Figure~\ref{hel-FP-para}(f). As for the single crystal diffraction data, the presence of these satellites around a structural Bragg reflection indicates the presence of a long-range periodically modulated spin texture. The data in the H phase was best fitted by a helical model, where the relative orientation of spins within the unit cell is fixed, but helically modulated between unit cells with a $Q$ of $0.0015$ \AA$^{-1}$. This corresponds to a helical state consisting of ferrimagnetic clusters, where we fit an average moment size of $0.74(4)$ $\mu_{\text{B}}$/Cu, and thus lower than the moment in the ferrimagnetic phase. We attribute this to the degenerate ground state, where the helical wavevector can point in any $\{100\}$ direction. Fluctuations can thus lead to a reduction in the observed magnetic moment.
 
\begin{figure}[b]
\centering
\includegraphics[width=\columnwidth]{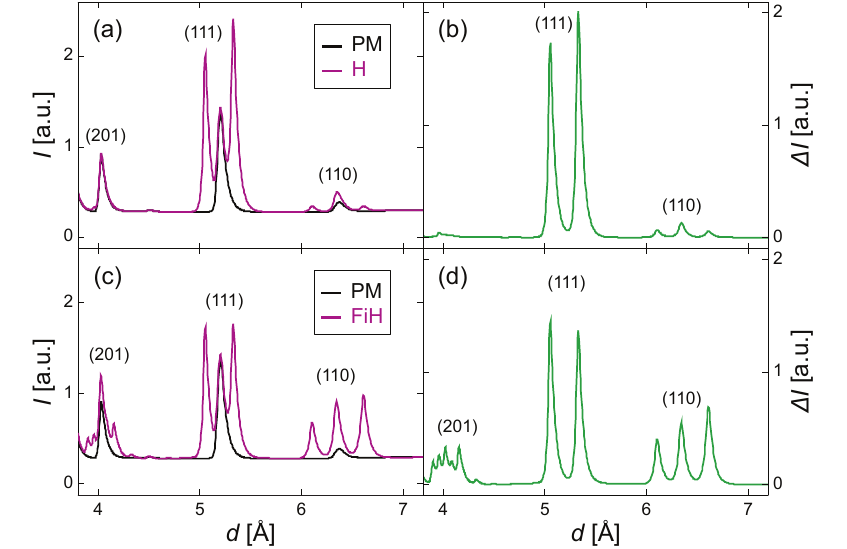}
\caption{\label{models}Simulations for powder patterns with exaggerated magnetic moments and helical wavevector of the PM (black lines) and H (violet) phases of Cu$_{2}$OSeO$_{3}$ for (a) helices made from individual spins, and (c) helices composed of ferrimagnetic clusters. The difference between simulated H and PM diffraction patterns are displayed in (b) and (d).}
\end{figure}
While neutron powder diffraction patterns can be fitted by Rietveld refinement, a qualitative analysis is instructive in understanding the helical ground state of Cu$_{2}$OSeO$_{3}$. 
To that end Figure~\ref{models} presents simulated powder diffraction patterns for (a) helices composed of individual spins, and (c) helices with ferrimagnetic alignment of spins within Cu tetrahedra. 
An exaggerated magnetic moment ($4$ $\mu_{\text{B}}$/Cu) and $Q$ ($0.08$ \AA$^{-1}$) have been used to better distinguish helical satellite peaks. 
Simulations reproducing the PM phase are included for comparison and the difference between patterns shown in Figures~\ref{models}(b) \& (d). Both models predict the absence of a magnetic $(111)$ peak, but the presence of magnetic satellites around it. They furthermore show a magnetic peak and satellites at $(110)$, however these are more pronounced in the ferrimagnetic model. Finally, only the ferrimagnetic model indicates a magnetic peak at the $(201)$ Bragg peak position. Comparing the neutron diffraction data presented in Figure~\ref{hel-FP-para} with the simulations in Figure~\ref{models} it becomes apparent, that the magnetic ground state of Cu$_{2}$OSeO$_{3}$ does not only consist of helices, but that these are made up of ferrimagnetically aligned clusters of spins: the H satellites predicted around the structural $(111)$ peak are clearly observed in Figure~\ref{hel-FP-para}(f). Magnetic satellites are not resolved around the (110) peak in the diffraction measurements, but a large magnetic peak is most likely consistent with the ferrimagnetic cluster model of the H phase. Finally, the magnetic peak observed at the $(201)$ position is only accounted for by the ferrimagnetic cluster model. We have thus established the magnetic ground state of Cu$_{2}$OSeO$_{3}$ to be composed of helices made from ferrimagnetically aligned spin clusters instead of individual spins.

\section{Conclusion}
Single crystal and powder long-wavelength neutron diffraction data was measured on the skyrmion host Cu$_{2}$OSeO$_{3}$. We observe magnetic satellites around the $(0\bar{1}1)$ diffraction peak not accessible by other techniques, and distinguish helical from conical spin textures in reciprocal space. Measurements show successive transitions from a helical to a conical to a field polarised magnetic phase as the external magnetic field is increased. While the resolution is too low to resolve the individual 6-fold skyrmion satellites, the formation of a skyrmion lattice with propagation vectors perpendicular to the field direction is observed.
As our key result, we show that not only the field polarised phase but also the helical ground state are made up of ferrimagnetic clusters instead of individual spins. These clusters are distorted Cu tetrahedra, where the spin on the Cu1 ion is anti-aligned with the spin on three Cu2 ions. Supporting magnetometry in the paramagnetic phase confirms the proposed ferrimagnetism.

\begin{acknowledgments}
We are grateful for the provision of beamtime at the Science and Technology Facilities Council (STFC) ISIS Facility, Rutherford Appleton Laboratory, UK. We acknowledge the use of the MPMS on the I10 beamline of the Diamond Light Source. 
This work was supported by the EPSRC through grants EP/M028771/1 and EP/N032128/1 and a scholarship for P R D. 
Research data will be made available via Durham Collections.
\end{acknowledgments}


\begin{thebibliography}{27}%
\makeatletter
\providecommand \@ifxundefined [1]{%
 \@ifx{#1\undefined}
}%
\providecommand \@ifnum [1]{%
 \ifnum #1\expandafter \@firstoftwo
 \else \expandafter \@secondoftwo
 \fi
}%
\providecommand \@ifx [1]{%
 \ifx #1\expandafter \@firstoftwo
 \else \expandafter \@secondoftwo
 \fi
}%
\providecommand \natexlab [1]{#1}%
\providecommand \enquote  [1]{``#1''}%
\providecommand \bibnamefont  [1]{#1}%
\providecommand \bibfnamefont [1]{#1}%
\providecommand \citenamefont [1]{#1}%
\providecommand \href@noop [0]{\@secondoftwo}%
\providecommand \href [0]{\begingroup \@sanitize@url \@href}%
\providecommand \@href[1]{\@@startlink{#1}\@@href}%
\providecommand \@@href[1]{\endgroup#1\@@endlink}%
\providecommand \@sanitize@url [0]{\catcode `\\12\catcode `\$12\catcode
  `\&12\catcode `\#12\catcode `\^12\catcode `\_12\catcode `\%12\relax}%
\providecommand \@@startlink[1]{}%
\providecommand \@@endlink[0]{}%
\providecommand \url  [0]{\begingroup\@sanitize@url \@url }%
\providecommand \@url [1]{\endgroup\@href {#1}{\urlprefix }}%
\providecommand \urlprefix  [0]{URL }%
\providecommand \Eprint [0]{\href }%
\providecommand \doibase [0]{http://dx.doi.org/}%
\providecommand \selectlanguage [0]{\@gobble}%
\providecommand \bibinfo  [0]{\@secondoftwo}%
\providecommand \bibfield  [0]{\@secondoftwo}%
\providecommand \translation [1]{[#1]}%
\providecommand \BibitemOpen [0]{}%
\providecommand \bibitemStop [0]{}%
\providecommand \bibitemNoStop [0]{.\EOS\space}%
\providecommand \EOS [0]{\spacefactor3000\relax}%
\providecommand \BibitemShut  [1]{\csname bibitem#1\endcsname}%
\let\auto@bib@innerbib\@empty
\bibitem [{\citenamefont {M\"uhlbauer}\ \emph {et~al.}(2009)\citenamefont
  {M\"uhlbauer}, \citenamefont {Binz}, \citenamefont {Jonietz}, \citenamefont
  {Pfleiderer}, \citenamefont {Rosch}, \citenamefont {Neubauer}, \citenamefont
  {Georgii},\ and\ \citenamefont {B\"oni}}]{muhlbauer_skyrmion_2009}%
  \BibitemOpen
  \bibfield  {author} {\bibinfo {author} {\bibfnamefont {S.}~\bibnamefont
  {M\"uhlbauer}}, \bibinfo {author} {\bibfnamefont {B.}~\bibnamefont {Binz}},
  \bibinfo {author} {\bibfnamefont {F.}~\bibnamefont {Jonietz}}, \bibinfo
  {author} {\bibfnamefont {C.}~\bibnamefont {Pfleiderer}}, \bibinfo {author}
  {\bibfnamefont {A.}~\bibnamefont {Rosch}}, \bibinfo {author} {\bibfnamefont
  {A.}~\bibnamefont {Neubauer}}, \bibinfo {author} {\bibfnamefont
  {R.}~\bibnamefont {Georgii}}, \ and\ \bibinfo {author} {\bibfnamefont
  {P.}~\bibnamefont {B\"oni}},\ }\href {\doibase 10.1126/science.1166767}
  {\bibfield  {journal} {\bibinfo  {journal} {Science}\ }\textbf {\bibinfo
  {volume} {323}},\ \bibinfo {pages} {915} (\bibinfo {year}
  {2009})}\BibitemShut {NoStop}%
\bibitem [{\citenamefont {Yu}\ \emph {et~al.}(2010)\citenamefont {Yu},
  \citenamefont {Onose}, \citenamefont {Kanazawa}, \citenamefont {Park},
  \citenamefont {Han}, \citenamefont {Matsui}, \citenamefont {Nagaosa},\ and\
  \citenamefont {Tokura}}]{yu_real-space_2010}%
  \BibitemOpen
  \bibfield  {author} {\bibinfo {author} {\bibfnamefont {X.~Z.}\ \bibnamefont
  {Yu}}, \bibinfo {author} {\bibfnamefont {Y.}~\bibnamefont {Onose}}, \bibinfo
  {author} {\bibfnamefont {N.}~\bibnamefont {Kanazawa}}, \bibinfo {author}
  {\bibfnamefont {J.~H.}\ \bibnamefont {Park}}, \bibinfo {author}
  {\bibfnamefont {J.~H.}\ \bibnamefont {Han}}, \bibinfo {author} {\bibfnamefont
  {Y.}~\bibnamefont {Matsui}}, \bibinfo {author} {\bibfnamefont
  {N.}~\bibnamefont {Nagaosa}}, \ and\ \bibinfo {author} {\bibfnamefont
  {Y.}~\bibnamefont {Tokura}},\ }\href {\doibase 10.1038/nature09124}
  {\bibfield  {journal} {\bibinfo  {journal} {Nature}\ }\textbf {\bibinfo
  {volume} {465}},\ \bibinfo {pages} {901} (\bibinfo {year}
  {2010})}\BibitemShut {NoStop}%
\bibitem [{\citenamefont {Yu}\ \emph {et~al.}(2011)\citenamefont {Yu},
  \citenamefont {Kanazawa}, \citenamefont {Onose}, \citenamefont {Kimoto},
  \citenamefont {Zhang}, \citenamefont {Ishiwata}, \citenamefont {Matsui},\
  and\ \citenamefont {Tokura}}]{yu_near_2011}%
  \BibitemOpen
  \bibfield  {author} {\bibinfo {author} {\bibfnamefont {X.~Z.}\ \bibnamefont
  {Yu}}, \bibinfo {author} {\bibfnamefont {N.}~\bibnamefont {Kanazawa}},
  \bibinfo {author} {\bibfnamefont {Y.}~\bibnamefont {Onose}}, \bibinfo
  {author} {\bibfnamefont {K.}~\bibnamefont {Kimoto}}, \bibinfo {author}
  {\bibfnamefont {W.~Z.}\ \bibnamefont {Zhang}}, \bibinfo {author}
  {\bibfnamefont {S.}~\bibnamefont {Ishiwata}}, \bibinfo {author}
  {\bibfnamefont {Y.}~\bibnamefont {Matsui}}, \ and\ \bibinfo {author}
  {\bibfnamefont {Y.}~\bibnamefont {Tokura}},\ }\href {\doibase
  10.1038/nmat2916} {\bibfield  {journal} {\bibinfo  {journal} {Nat. Mater.}\
  }\textbf {\bibinfo {volume} {10}},\ \bibinfo {pages} {106} (\bibinfo {year}
  {2011})}\BibitemShut {NoStop}%
\bibitem [{\citenamefont {Seki}\ \emph
  {et~al.}(2012{\natexlab{a}})\citenamefont {Seki}, \citenamefont {Yu},
  \citenamefont {Ishiwata},\ and\ \citenamefont
  {Tokura}}]{seki_observation_2012}%
  \BibitemOpen
  \bibfield  {author} {\bibinfo {author} {\bibfnamefont {S.}~\bibnamefont
  {Seki}}, \bibinfo {author} {\bibfnamefont {X.~Z.}\ \bibnamefont {Yu}},
  \bibinfo {author} {\bibfnamefont {S.}~\bibnamefont {Ishiwata}}, \ and\
  \bibinfo {author} {\bibfnamefont {Y.}~\bibnamefont {Tokura}},\ }\href
  {\doibase 10.1126/science.1214143} {\bibfield  {journal} {\bibinfo  {journal}
  {Science}\ }\textbf {\bibinfo {volume} {336}},\ \bibinfo {pages} {198}
  (\bibinfo {year} {2012}{\natexlab{a}})}\BibitemShut {NoStop}%
\bibitem [{\citenamefont {Tokunaga}\ \emph {et~al.}(2015)\citenamefont
  {Tokunaga}, \citenamefont {Yu}, \citenamefont {White}, \citenamefont
  {R{\o}nnow}, \citenamefont {Morikawa}, \citenamefont {Taguchi},\ and\
  \citenamefont {Tokura}}]{tokunaga_new_2015}%
  \BibitemOpen
  \bibfield  {author} {\bibinfo {author} {\bibfnamefont {Y.}~\bibnamefont
  {Tokunaga}}, \bibinfo {author} {\bibfnamefont {X.~Z.}\ \bibnamefont {Yu}},
  \bibinfo {author} {\bibfnamefont {J.~S.}\ \bibnamefont {White}}, \bibinfo
  {author} {\bibfnamefont {H.~M.}\ \bibnamefont {R{\o}nnow}}, \bibinfo {author}
  {\bibfnamefont {D.}~\bibnamefont {Morikawa}}, \bibinfo {author}
  {\bibfnamefont {Y.}~\bibnamefont {Taguchi}}, \ and\ \bibinfo {author}
  {\bibfnamefont {Y.}~\bibnamefont {Tokura}},\ }\href {\doibase
  10.1038/ncomms8638} {\bibfield  {journal} {\bibinfo  {journal} {Nat.
  Commun.}\ }\textbf {\bibinfo {volume} {6}},\ \bibinfo {pages} {ncomms8638}
  (\bibinfo {year} {2015})}\BibitemShut {NoStop}%
\bibitem [{\citenamefont {K\'ezsm\'arki}\ \emph {et~al.}(2015)\citenamefont
  {K\'ezsm\'arki}, \citenamefont {Bord\'acs}, \citenamefont {Milde},
  \citenamefont {Neuber}, \citenamefont {Eng}, \citenamefont {White},
  \citenamefont {R{\o}nnow}, \citenamefont {Dewhurst}, \citenamefont
  {Mochizuki}, \citenamefont {Yanai}, \citenamefont {Nakamura}, \citenamefont
  {Ehlers}, \citenamefont {Tsurkan},\ and\ \citenamefont
  {Loidl}}]{kezsmarki_neel-type_2015}%
  \BibitemOpen
  \bibfield  {author} {\bibinfo {author} {\bibfnamefont {I.}~\bibnamefont
  {K\'ezsm\'arki}}, \bibinfo {author} {\bibfnamefont {S.}~\bibnamefont
  {Bord\'acs}}, \bibinfo {author} {\bibfnamefont {P.}~\bibnamefont {Milde}},
  \bibinfo {author} {\bibfnamefont {E.}~\bibnamefont {Neuber}}, \bibinfo
  {author} {\bibfnamefont {L.~M.}\ \bibnamefont {Eng}}, \bibinfo {author}
  {\bibfnamefont {J.~S.}\ \bibnamefont {White}}, \bibinfo {author}
  {\bibfnamefont {H.~M.}\ \bibnamefont {R{\o}nnow}}, \bibinfo {author}
  {\bibfnamefont {C.~D.}\ \bibnamefont {Dewhurst}}, \bibinfo {author}
  {\bibfnamefont {M.}~\bibnamefont {Mochizuki}}, \bibinfo {author}
  {\bibfnamefont {K.}~\bibnamefont {Yanai}}, \bibinfo {author} {\bibfnamefont
  {H.}~\bibnamefont {Nakamura}}, \bibinfo {author} {\bibfnamefont
  {D.}~\bibnamefont {Ehlers}}, \bibinfo {author} {\bibfnamefont
  {V.}~\bibnamefont {Tsurkan}}, \ and\ \bibinfo {author} {\bibfnamefont
  {A.}~\bibnamefont {Loidl}},\ }\href {\doibase 10.1038/nmat4402} {\bibfield
  {journal} {\bibinfo  {journal} {Nat. Mater.}\ }\textbf {\bibinfo {volume}
  {14}},\ \bibinfo {pages} {1116} (\bibinfo {year} {2015})}\BibitemShut
  {NoStop}%
\bibitem [{\citenamefont {Bord\'acs}\ \emph {et~al.}(2017)\citenamefont
  {Bord\'acs}, \citenamefont {Butykai}, \citenamefont {Szigeti}, \citenamefont
  {White}, \citenamefont {Cubitt}, \citenamefont {Leonov}, \citenamefont
  {Widmann}, \citenamefont {Ehlers}, \citenamefont {Nidda}, \citenamefont
  {Tsurkan}, \citenamefont {Loidl},\ and\ \citenamefont
  {K\'ezsm\'arki}}]{bordacs_equilibrium_2017}%
  \BibitemOpen
  \bibfield  {author} {\bibinfo {author} {\bibfnamefont {S.}~\bibnamefont
  {Bord\'acs}}, \bibinfo {author} {\bibfnamefont {A.}~\bibnamefont {Butykai}},
  \bibinfo {author} {\bibfnamefont {B.~G.}\ \bibnamefont {Szigeti}}, \bibinfo
  {author} {\bibfnamefont {J.~S.}\ \bibnamefont {White}}, \bibinfo {author}
  {\bibfnamefont {R.}~\bibnamefont {Cubitt}}, \bibinfo {author} {\bibfnamefont
  {A.~O.}\ \bibnamefont {Leonov}}, \bibinfo {author} {\bibfnamefont
  {S.}~\bibnamefont {Widmann}}, \bibinfo {author} {\bibfnamefont
  {D.}~\bibnamefont {Ehlers}}, \bibinfo {author} {\bibfnamefont {H.-A.~K.}\
  \bibnamefont {Nidda}}, \bibinfo {author} {\bibfnamefont {V.}~\bibnamefont
  {Tsurkan}}, \bibinfo {author} {\bibfnamefont {A.}~\bibnamefont {Loidl}}, \
  and\ \bibinfo {author} {\bibfnamefont {I.}~\bibnamefont {K\'ezsm\'arki}},\
  }\href {\doibase 10.1038/s41598-017-07996-x} {\bibfield  {journal} {\bibinfo
  {journal} {Sci. Rep.}\ }\textbf {\bibinfo {volume} {7}},\ \bibinfo {pages}
  {7584} (\bibinfo {year} {2017})}\BibitemShut {NoStop}%
\bibitem [{\citenamefont {Kurumaji}\ \emph {et~al.}(2017)\citenamefont
  {Kurumaji}, \citenamefont {Nakajima}, \citenamefont {Ukleev}, \citenamefont
  {Feoktystov}, \citenamefont {Arima}, \citenamefont {Kakurai},\ and\
  \citenamefont {Tokura}}]{kurumaji_neel-type_2017}%
  \BibitemOpen
  \bibfield  {author} {\bibinfo {author} {\bibfnamefont {T.}~\bibnamefont
  {Kurumaji}}, \bibinfo {author} {\bibfnamefont {T.}~\bibnamefont {Nakajima}},
  \bibinfo {author} {\bibfnamefont {V.}~\bibnamefont {Ukleev}}, \bibinfo
  {author} {\bibfnamefont {A.}~\bibnamefont {Feoktystov}}, \bibinfo {author}
  {\bibfnamefont {T.-h.}\ \bibnamefont {Arima}}, \bibinfo {author}
  {\bibfnamefont {K.}~\bibnamefont {Kakurai}}, \ and\ \bibinfo {author}
  {\bibfnamefont {Y.}~\bibnamefont {Tokura}},\ }\href {\doibase
  10.1103/PhysRevLett.119.237201} {\bibfield  {journal} {\bibinfo  {journal}
  {Phys. Rev. Lett.}\ }\textbf {\bibinfo {volume} {119}},\ \bibinfo {pages}
  {237201} (\bibinfo {year} {2017})}\BibitemShut {NoStop}%
\bibitem [{\citenamefont
  {Dzyaloshinsky}(1958)}]{dzyaloshinsky_thermodynamic_1958}%
  \BibitemOpen
  \bibfield  {author} {\bibinfo {author} {\bibfnamefont {I.}~\bibnamefont
  {Dzyaloshinsky}},\ }\href {\doibase 10.1016/0022-3697(58)90076-3} {\bibfield
  {journal} {\bibinfo  {journal} {J. Phys. Chem. Solids}\ }\textbf {\bibinfo
  {volume} {4}},\ \bibinfo {pages} {241} (\bibinfo {year} {1958})}\BibitemShut
  {NoStop}%
\bibitem [{\citenamefont {Moriya}(1960)}]{moriya_anisotropic_1960}%
  \BibitemOpen
  \bibfield  {author} {\bibinfo {author} {\bibfnamefont {T.}~\bibnamefont
  {Moriya}},\ }\href {\doibase 10.1103/PhysRev.120.91} {\bibfield  {journal}
  {\bibinfo  {journal} {Phys. Rev.}\ }\textbf {\bibinfo {volume} {120}},\
  \bibinfo {pages} {91} (\bibinfo {year} {1960})}\BibitemShut {NoStop}%
\bibitem [{\citenamefont {{A. N. Bogdanov}}\ and\ \citenamefont
  {Yablonskii}(1989)}]{bogdanov_thermodynamically_1989}%
  \BibitemOpen
  \bibfield  {author} {\bibinfo {author} {\bibnamefont {{A. N. Bogdanov}}}\
  and\ \bibinfo {author} {\bibfnamefont {D.~A.}\ \bibnamefont {Yablonskii}},\
  }\href@noop {} {\bibfield  {journal} {\bibinfo  {journal} {Sov. Phys. JETP}\
  }\textbf {\bibinfo {volume} {68}},\ \bibinfo {pages} {101} (\bibinfo {year}
  {1989})}\BibitemShut {NoStop}%
\bibitem [{\citenamefont {Bogdanov}\ and\ \citenamefont
  {Hubert}(1994)}]{bogdanov_thermodynamically_1994}%
  \BibitemOpen
  \bibfield  {author} {\bibinfo {author} {\bibfnamefont {A.}~\bibnamefont
  {Bogdanov}}\ and\ \bibinfo {author} {\bibfnamefont {A.}~\bibnamefont
  {Hubert}},\ }\href {\doibase 10.1016/0304-8853(94)90046-9} {\bibfield
  {journal} {\bibinfo  {journal} {Journal of Magnetism and Magnetic Materials}\
  }\textbf {\bibinfo {volume} {138}},\ \bibinfo {pages} {255} (\bibinfo {year}
  {1994})}\BibitemShut {NoStop}%
\bibitem [{\citenamefont {Adams}\ \emph {et~al.}(2012)\citenamefont {Adams},
  \citenamefont {Chacon}, \citenamefont {Wagner}, \citenamefont {Bauer},
  \citenamefont {Brandl}, \citenamefont {Pedersen}, \citenamefont {Berger},
  \citenamefont {Lemmens},\ and\ \citenamefont
  {Pfleiderer}}]{adams_long-wavelength_2012}%
  \BibitemOpen
  \bibfield  {author} {\bibinfo {author} {\bibfnamefont {T.}~\bibnamefont
  {Adams}}, \bibinfo {author} {\bibfnamefont {A.}~\bibnamefont {Chacon}},
  \bibinfo {author} {\bibfnamefont {M.}~\bibnamefont {Wagner}}, \bibinfo
  {author} {\bibfnamefont {A.}~\bibnamefont {Bauer}}, \bibinfo {author}
  {\bibfnamefont {G.}~\bibnamefont {Brandl}}, \bibinfo {author} {\bibfnamefont
  {B.}~\bibnamefont {Pedersen}}, \bibinfo {author} {\bibfnamefont
  {H.}~\bibnamefont {Berger}}, \bibinfo {author} {\bibfnamefont
  {P.}~\bibnamefont {Lemmens}}, \ and\ \bibinfo {author} {\bibfnamefont
  {C.}~\bibnamefont {Pfleiderer}},\ }\href {\doibase
  10.1103/PhysRevLett.108.237204} {\bibfield  {journal} {\bibinfo  {journal}
  {Phys. Rev. Lett.}\ }\textbf {\bibinfo {volume} {108}},\ \bibinfo {pages}
  {237204} (\bibinfo {year} {2012})}\BibitemShut {NoStop}%
\bibitem [{\citenamefont {Seki}\ \emph
  {et~al.}(2012{\natexlab{b}})\citenamefont {Seki}, \citenamefont {Kim},
  \citenamefont {Inosov}, \citenamefont {Georgii}, \citenamefont {Keimer},
  \citenamefont {Ishiwata},\ and\ \citenamefont
  {Tokura}}]{seki_formation_2012}%
  \BibitemOpen
  \bibfield  {author} {\bibinfo {author} {\bibfnamefont {S.}~\bibnamefont
  {Seki}}, \bibinfo {author} {\bibfnamefont {J.-H.}\ \bibnamefont {Kim}},
  \bibinfo {author} {\bibfnamefont {D.~S.}\ \bibnamefont {Inosov}}, \bibinfo
  {author} {\bibfnamefont {R.}~\bibnamefont {Georgii}}, \bibinfo {author}
  {\bibfnamefont {B.}~\bibnamefont {Keimer}}, \bibinfo {author} {\bibfnamefont
  {S.}~\bibnamefont {Ishiwata}}, \ and\ \bibinfo {author} {\bibfnamefont
  {Y.}~\bibnamefont {Tokura}},\ }\href {\doibase 10.1103/PhysRevB.85.220406}
  {\bibfield  {journal} {\bibinfo  {journal} {Phys. Rev. B}\ }\textbf {\bibinfo
  {volume} {85}},\ \bibinfo {pages} {220406} (\bibinfo {year}
  {2012}{\natexlab{b}})}\BibitemShut {NoStop}%
\bibitem [{\citenamefont {Yang}\ \emph {et~al.}(2012)\citenamefont {Yang},
  \citenamefont {Li}, \citenamefont {Lu}, \citenamefont {Whangbo},
  \citenamefont {Wei}, \citenamefont {Gong},\ and\ \citenamefont
  {Xiang}}]{yang_strong_2012}%
  \BibitemOpen
  \bibfield  {author} {\bibinfo {author} {\bibfnamefont {J.~H.}\ \bibnamefont
  {Yang}}, \bibinfo {author} {\bibfnamefont {Z.~L.}\ \bibnamefont {Li}},
  \bibinfo {author} {\bibfnamefont {X.~Z.}\ \bibnamefont {Lu}}, \bibinfo
  {author} {\bibfnamefont {M.-H.}\ \bibnamefont {Whangbo}}, \bibinfo {author}
  {\bibfnamefont {S.-H.}\ \bibnamefont {Wei}}, \bibinfo {author} {\bibfnamefont
  {X.~G.}\ \bibnamefont {Gong}}, \ and\ \bibinfo {author} {\bibfnamefont
  {H.~J.}\ \bibnamefont {Xiang}},\ }\href {\doibase
  10.1103/PhysRevLett.109.107203} {\bibfield  {journal} {\bibinfo  {journal}
  {Phys. Rev. Lett.}\ }\textbf {\bibinfo {volume} {109}},\ \bibinfo {pages}
  {107203} (\bibinfo {year} {2012})}\BibitemShut {NoStop}%
\bibitem [{\citenamefont {Bos}\ \emph {et~al.}(2008)\citenamefont {Bos},
  \citenamefont {Colin},\ and\ \citenamefont
  {Palstra}}]{bos_magnetoelectric_2008}%
  \BibitemOpen
  \bibfield  {author} {\bibinfo {author} {\bibfnamefont {J.-W.~G.}\
  \bibnamefont {Bos}}, \bibinfo {author} {\bibfnamefont {C.~V.}\ \bibnamefont
  {Colin}}, \ and\ \bibinfo {author} {\bibfnamefont {T.~T.~M.}\ \bibnamefont
  {Palstra}},\ }\href {\doibase 10.1103/PhysRevB.78.094416} {\bibfield
  {journal} {\bibinfo  {journal} {Phys. Rev. B}\ }\textbf {\bibinfo {volume}
  {78}},\ \bibinfo {pages} {094416} (\bibinfo {year} {2008})}\BibitemShut
  {NoStop}%
\bibitem [{\citenamefont {Kohn}(1977)}]{kohn_new_1977}%
  \BibitemOpen
  \bibfield  {author} {\bibinfo {author} {\bibfnamefont {K.}~\bibnamefont
  {Kohn}},\ }\href {\doibase 10.1143/JPSJ.42.2065} {\bibfield  {journal}
  {\bibinfo  {journal} {J. Phys. Soc. Jpn.}\ }\textbf {\bibinfo {volume}
  {42}},\ \bibinfo {pages} {2065} (\bibinfo {year} {1977})}\BibitemShut
  {NoStop}%
\bibitem [{\citenamefont {Belesi}\ \emph {et~al.}(2010)\citenamefont {Belesi},
  \citenamefont {Rousochatzakis}, \citenamefont {Wu}, \citenamefont {Berger},
  \citenamefont {Shvets}, \citenamefont {Mila},\ and\ \citenamefont
  {Ansermet}}]{belesi_ferrimagnetism_2010}%
  \BibitemOpen
  \bibfield  {author} {\bibinfo {author} {\bibfnamefont {M.}~\bibnamefont
  {Belesi}}, \bibinfo {author} {\bibfnamefont {I.}~\bibnamefont
  {Rousochatzakis}}, \bibinfo {author} {\bibfnamefont {H.~C.}\ \bibnamefont
  {Wu}}, \bibinfo {author} {\bibfnamefont {H.}~\bibnamefont {Berger}}, \bibinfo
  {author} {\bibfnamefont {I.~V.}\ \bibnamefont {Shvets}}, \bibinfo {author}
  {\bibfnamefont {F.}~\bibnamefont {Mila}}, \ and\ \bibinfo {author}
  {\bibfnamefont {J.~P.}\ \bibnamefont {Ansermet}},\ }\href {\doibase
  10.1103/PhysRevB.82.094422} {\bibfield  {journal} {\bibinfo  {journal} {Phys.
  Rev. B}\ }\textbf {\bibinfo {volume} {82}},\ \bibinfo {pages} {094422}
  (\bibinfo {year} {2010})}\BibitemShut {NoStop}%
\bibitem [{\citenamefont {Zhang}\ \emph {et~al.}(2017)\citenamefont {Zhang},
  \citenamefont {Laan},\ and\ \citenamefont {Hesjedal}}]{zhang_direct_2017}%
  \BibitemOpen
  \bibfield  {author} {\bibinfo {author} {\bibfnamefont {S.~L.}\ \bibnamefont
  {Zhang}}, \bibinfo {author} {\bibfnamefont {G.~v.~d.}\ \bibnamefont {Laan}},
  \ and\ \bibinfo {author} {\bibfnamefont {T.}~\bibnamefont {Hesjedal}},\
  }\href {\doibase 10.1038/ncomms14619} {\bibfield  {journal} {\bibinfo
  {journal} {Nature Communications}\ }\textbf {\bibinfo {volume} {8}},\
  \bibinfo {pages} {14619} (\bibinfo {year} {2017})}\BibitemShut {NoStop}%
\bibitem [{\citenamefont {Chapon}\ \emph {et~al.}(2011)\citenamefont {Chapon},
  \citenamefont {Manuel}, \citenamefont {Radaelli}, \citenamefont {Benson},
  \citenamefont {Perrott}, \citenamefont {Ansell}, \citenamefont {Rhodes},
  \citenamefont {Raspino}, \citenamefont {Duxbury}, \citenamefont {Spill},\
  and\ \citenamefont {Norris}}]{chapon_wish:_2011}%
  \BibitemOpen
  \bibfield  {author} {\bibinfo {author} {\bibfnamefont {L.~C.}\ \bibnamefont
  {Chapon}}, \bibinfo {author} {\bibfnamefont {P.}~\bibnamefont {Manuel}},
  \bibinfo {author} {\bibfnamefont {P.~G.}\ \bibnamefont {Radaelli}}, \bibinfo
  {author} {\bibfnamefont {C.}~\bibnamefont {Benson}}, \bibinfo {author}
  {\bibfnamefont {L.}~\bibnamefont {Perrott}}, \bibinfo {author} {\bibfnamefont
  {S.}~\bibnamefont {Ansell}}, \bibinfo {author} {\bibfnamefont {N.~J.}\
  \bibnamefont {Rhodes}}, \bibinfo {author} {\bibfnamefont {D.}~\bibnamefont
  {Raspino}}, \bibinfo {author} {\bibfnamefont {D.}~\bibnamefont {Duxbury}},
  \bibinfo {author} {\bibfnamefont {E.}~\bibnamefont {Spill}}, \ and\ \bibinfo
  {author} {\bibfnamefont {J.}~\bibnamefont {Norris}},\ }\href {\doibase
  10.1080/10448632.2011.569650} {\bibfield  {journal} {\bibinfo  {journal}
  {Neutron News}\ }\textbf {\bibinfo {volume} {22}},\ \bibinfo {pages} {22}
  (\bibinfo {year} {2011})}\BibitemShut {NoStop}%
\bibitem [{\citenamefont {Arnold}\ \emph {et~al.}(2014)\citenamefont {Arnold},
  \citenamefont {Bilheux}, \citenamefont {Borreguero}, \citenamefont {Buts},
  \citenamefont {Campbell}, \citenamefont {Chapon}, \citenamefont {Doucet},
  \citenamefont {Draper}, \citenamefont {Leal}, \citenamefont {Gigg},
  \citenamefont {Lynch}, \citenamefont {Markvardsen}, \citenamefont
  {Mikkelson}, \citenamefont {Mikkelson}, \citenamefont {Miller}, \citenamefont
  {Palmen}, \citenamefont {Parker}, \citenamefont {Passos}, \citenamefont
  {Perring}, \citenamefont {Peterson}, \citenamefont {Ren}, \citenamefont
  {Reuter}, \citenamefont {Savici}, \citenamefont {Taylor}, \citenamefont
  {Taylor}, \citenamefont {Tolchenov}, \citenamefont {Zhou},\ and\
  \citenamefont {Zikovsky}}]{Mantid2014}%
  \BibitemOpen
  \bibfield  {author} {\bibinfo {author} {\bibfnamefont {O.}~\bibnamefont
  {Arnold}}, \bibinfo {author} {\bibfnamefont {J.}~\bibnamefont {Bilheux}},
  \bibinfo {author} {\bibfnamefont {J.}~\bibnamefont {Borreguero}}, \bibinfo
  {author} {\bibfnamefont {A.}~\bibnamefont {Buts}}, \bibinfo {author}
  {\bibfnamefont {S.}~\bibnamefont {Campbell}}, \bibinfo {author}
  {\bibfnamefont {L.}~\bibnamefont {Chapon}}, \bibinfo {author} {\bibfnamefont
  {M.}~\bibnamefont {Doucet}}, \bibinfo {author} {\bibfnamefont
  {N.}~\bibnamefont {Draper}}, \bibinfo {author} {\bibfnamefont {R.~F.}\
  \bibnamefont {Leal}}, \bibinfo {author} {\bibfnamefont {M.}~\bibnamefont
  {Gigg}}, \bibinfo {author} {\bibfnamefont {V.}~\bibnamefont {Lynch}},
  \bibinfo {author} {\bibfnamefont {A.}~\bibnamefont {Markvardsen}}, \bibinfo
  {author} {\bibfnamefont {D.}~\bibnamefont {Mikkelson}}, \bibinfo {author}
  {\bibfnamefont {R.}~\bibnamefont {Mikkelson}}, \bibinfo {author}
  {\bibfnamefont {R.}~\bibnamefont {Miller}}, \bibinfo {author} {\bibfnamefont
  {K.}~\bibnamefont {Palmen}}, \bibinfo {author} {\bibfnamefont
  {P.}~\bibnamefont {Parker}}, \bibinfo {author} {\bibfnamefont
  {G.}~\bibnamefont {Passos}}, \bibinfo {author} {\bibfnamefont
  {T.}~\bibnamefont {Perring}}, \bibinfo {author} {\bibfnamefont
  {P.}~\bibnamefont {Peterson}}, \bibinfo {author} {\bibfnamefont
  {S.}~\bibnamefont {Ren}}, \bibinfo {author} {\bibfnamefont {M.}~\bibnamefont
  {Reuter}}, \bibinfo {author} {\bibfnamefont {A.}~\bibnamefont {Savici}},
  \bibinfo {author} {\bibfnamefont {J.}~\bibnamefont {Taylor}}, \bibinfo
  {author} {\bibfnamefont {R.}~\bibnamefont {Taylor}}, \bibinfo {author}
  {\bibfnamefont {R.}~\bibnamefont {Tolchenov}}, \bibinfo {author}
  {\bibfnamefont {W.}~\bibnamefont {Zhou}}, \ and\ \bibinfo {author}
  {\bibfnamefont {J.}~\bibnamefont {Zikovsky}},\ }\href {\doibase
  https://doi.org/10.1016/j.nima.2014.07.029} {\bibfield  {journal} {\bibinfo
  {journal} {Nucl. Instrum. Methods Phys. Res. A}\ }\textbf {\bibinfo {volume}
  {764}},\ \bibinfo {pages} {156 } (\bibinfo {year} {2014})}\BibitemShut
  {NoStop}%
\bibitem [{\citenamefont {Rodr\'iguez-Carvajal}(1993)}]{FullProf1993}%
  \BibitemOpen
  \bibfield  {author} {\bibinfo {author} {\bibfnamefont {J.}~\bibnamefont
  {Rodr\'iguez-Carvajal}},\ }\href {\doibase
  https://doi.org/10.1016/0921-4526(93)90108-I} {\bibfield  {journal} {\bibinfo
   {journal} {Physica B Condens. Matter}\ }\textbf {\bibinfo {volume} {192}},\
  \bibinfo {pages} {55 } (\bibinfo {year} {1993})}\BibitemShut {NoStop}%
\bibitem [{\citenamefont {Bauer}\ and\ \citenamefont
  {Pfleiderer}(2012)}]{bauer_magnetic_2012}%
  \BibitemOpen
  \bibfield  {author} {\bibinfo {author} {\bibfnamefont {A.}~\bibnamefont
  {Bauer}}\ and\ \bibinfo {author} {\bibfnamefont {C.}~\bibnamefont
  {Pfleiderer}},\ }\href {\doibase 10.1103/PhysRevB.85.214418} {\bibfield
  {journal} {\bibinfo  {journal} {Phys. Rev. B}\ }\textbf {\bibinfo {volume}
  {85}},\ \bibinfo {pages} {214418} (\bibinfo {year} {2012})}\BibitemShut
  {NoStop}%
\bibitem [{\citenamefont {Bauer}\ and\ \citenamefont
  {Pfleiderer}(2016)}]{bauer_generic_2016}%
  \BibitemOpen
  \bibfield  {author} {\bibinfo {author} {\bibfnamefont {A.}~\bibnamefont
  {Bauer}}\ and\ \bibinfo {author} {\bibfnamefont {C.}~\bibnamefont
  {Pfleiderer}},\ }in\ \href {\doibase 10.1007/978-3-319-25301-5_1} {\emph
  {\bibinfo {booktitle} {Topological {Structures} in {Ferroic} {Materials}}}},\
  Vol.\ \bibinfo {volume} {228},\ \bibinfo {editor} {edited by\ \bibinfo
  {editor} {\bibfnamefont {J.}~\bibnamefont {Seidel}}}\ (\bibinfo  {publisher}
  {Springer International Publishing},\ \bibinfo {address} {Cham},\ \bibinfo
  {year} {2016})\ pp.\ \bibinfo {pages} {1--28}\BibitemShut {NoStop}%
\bibitem [{\citenamefont {Kittel}(1996)}]{kittel_introduction_1996}%
  \BibitemOpen
  \bibfield  {author} {\bibinfo {author} {\bibfnamefont {C.}~\bibnamefont
  {Kittel}},\ }\href@noop {} {\emph {\bibinfo {title} {Introduction to {Solid}
  {State} {Physics}}}},\ \bibinfo {edition} {7th}\ ed.\ (\bibinfo  {publisher}
  {Wiley},\ \bibinfo {address} {Hoboken, NJ},\ \bibinfo {year}
  {1996})\BibitemShut {NoStop}%
\bibitem [{\citenamefont {Langner}\ \emph {et~al.}(2014)\citenamefont
  {Langner}, \citenamefont {Roy}, \citenamefont {Mishra}, \citenamefont {Lee},
  \citenamefont {Shi}, \citenamefont {Hossain}, \citenamefont {Chuang},
  \citenamefont {Seki}, \citenamefont {Tokura}, \citenamefont {Kevan},\ and\
  \citenamefont {Schoenlein}}]{langner_coupled_2014}%
  \BibitemOpen
  \bibfield  {author} {\bibinfo {author} {\bibfnamefont {M.~C.}\ \bibnamefont
  {Langner}}, \bibinfo {author} {\bibfnamefont {S.}~\bibnamefont {Roy}},
  \bibinfo {author} {\bibfnamefont {S.~K.}\ \bibnamefont {Mishra}}, \bibinfo
  {author} {\bibfnamefont {J.~C.~�.}\ \bibnamefont {Lee}}, \bibinfo {author}
  {\bibfnamefont {X.~�.}\ \bibnamefont {Shi}}, \bibinfo {author} {\bibfnamefont
  {M.~A.}\ \bibnamefont {Hossain}}, \bibinfo {author} {\bibfnamefont {Y.-D.}\
  \bibnamefont {Chuang}}, \bibinfo {author} {\bibfnamefont {S.}~\bibnamefont
  {Seki}}, \bibinfo {author} {\bibfnamefont {Y.}~\bibnamefont {Tokura}},
  \bibinfo {author} {\bibfnamefont {S.~�.}\ \bibnamefont {Kevan}}, \ and\
  \bibinfo {author} {\bibfnamefont {R.~�.}\ \bibnamefont {Schoenlein}},\ }\href
  {\doibase 10.1103/PhysRevLett.112.167202} {\bibfield  {journal} {\bibinfo
  {journal} {Phys. Rev. Lett.}\ }\textbf {\bibinfo {volume} {112}},\ \bibinfo
  {pages} {167202} (\bibinfo {year} {2014})}\BibitemShut {NoStop}%
\bibitem [{\citenamefont {Zhang}\ \emph {et~al.}(2016)\citenamefont {Zhang},
  \citenamefont {Bauer}, \citenamefont {Berger}, \citenamefont {Pfleiderer},
  \citenamefont {van~der Laan},\ and\ \citenamefont
  {Hesjedal}}]{zhang_resonant_2016}%
  \BibitemOpen
  \bibfield  {author} {\bibinfo {author} {\bibfnamefont {S.~L.}\ \bibnamefont
  {Zhang}}, \bibinfo {author} {\bibfnamefont {A.}~\bibnamefont {Bauer}},
  \bibinfo {author} {\bibfnamefont {H.}~\bibnamefont {Berger}}, \bibinfo
  {author} {\bibfnamefont {C.}~\bibnamefont {Pfleiderer}}, \bibinfo {author}
  {\bibfnamefont {G.}~\bibnamefont {van~der Laan}}, \ and\ \bibinfo {author}
  {\bibfnamefont {T.}~\bibnamefont {Hesjedal}},\ }\href {\doibase
  10.1103/PhysRevB.93.214420} {\bibfield  {journal} {\bibinfo  {journal} {Phys.
  Rev. B}\ }\textbf {\bibinfo {volume} {93}},\ \bibinfo {pages} {214420}
  (\bibinfo {year} {2016})}\BibitemShut {NoStop}%
\end{thebibliography}
\end{document}